\documentclass{article}
\usepackage{graphicx} % Required for inserting images
\usepackage{amsmath}
\usepackage{amsfonts}
\usepackage{hyperref}
\hypersetup{
    colorlinks=true,
    linkcolor=blue,
    filecolor=blue,
    citecolor = black,
    urlcolor=blue,
}
\usepackage{tikz}
\usepackage{pgfplots}
\usepackage{amssymb}
\usepackage{color,xcolor,soul}

\newcommand{\circlesign}[1]{ 
    \mathbin{
        \mathchoice {\buildcirclesign{\displaystyle}{#1}}{\buildcirclesign{\textstyle}{#1}}{\buildcirclesign{\scriptstyle}{#1}}{\buildcirclesign{\scriptscriptstyle}{#1}}
    } 
}
\newcommand\buildcirclesign[2]{%
    \begin{tikzpicture}[baseline=(X.base), inner sep=0, outer sep=0]
    \node[draw,circle] (X)  {\ensuremath{#1 #2}};
    \end{tikzpicture}%
}
\usepackage{tabularx}

\title{Post-Quantum Cryptography: An Analysis of Code-Based and Lattice-Based Cryptosystems}
\author{Alexander Meyer}
\date{August 2024}

\begin{document}
\maketitle

\section{Introduction} 
    Although not fully developed yet, quantum computers are expected to emerge in the near future, likely within ten to twenty years, which will pose a great threat to our encrypted data. Most modern cryptographic systems are not unbreakable but are presumed to be computationally difficult with the tools currently available. These systems are centered around the concept of “trapdoor” mathematical functions that work easily in one direction (encryption) and harder in the other (decryption) without access to a special key. Examples of these prominent cryptosystems include RSA, the Diffie-Hellman Key Exchange, and elliptic curve cryptography, all of which are based on problems that are presumed to be difficult and expensive for a classical computer to perform. However, quantum computers will be able to crack these systems rapidly using Shor's Algorithm, which is why research is currently being conducted into post-quantum cryptography to discover cryptographic methods that can withstand attacks from quantum computers \cite{Shor}.

    In 2016, the National Institute of Standards and Technology (\href{https://csrc.nist.gov/projects/post-quantum-cryptography}{NIST}) launched a competition to discover candidates for post-quantum cryptography, where experts submit proposals for new cryptosystems that will then undergo evaluation and attacks to determine their security. As the competition began, various kinds of cryptosystems were proposed and many were broken; however, lattice-based, code-based, and multivariate systems seemed to be the strongest. This paper will focus on both code-based and lattice-based systems and will also explore the connection between these two seemingly distinct cryptographic methods.

\section{Code-Based Cryptography}
Code-based cryptographic systems incorporate error-correcting, linear codes, and their security is typically based on the difficulty of finding the nearest codeword for an arbitrary linear code \cite{UC2024}. One of the most significant code-based systems is the McEliece cryptosystem, which uses a special class of codes, binary Goppa codes. These codes have an efficient decoding algorithm in Patterson's Algorithm, which allows the owner of the private key to decrypt a ciphertext quickly, whereas an eavesdropper cannot \cite{Patterson}.

\subsection{Binary Goppa Codes}
Let $g$ be an irreducible polynomial of degree $t$ over the finite field $GF(2^m)$ such that
\[g(x)=\sum^{t}_{i=0} g_ix^i=g_0+g_1x+...+g_ix^i\]
with $g_i\in GF(2^m)$ \cite{Repka}. Let $L$ be the sequence $L=(\alpha_0,..., \alpha_{n-1})$ of n distinct elements in $GF(2^m)$ such that $g(\alpha_i)\neq 0$, for $0\le i<n$ \cite{Repka}. The binary Goppa code $C(g,L)$ is defined as \[C(g,L)=\left\{c\in \mathbb{Z}^n_2\Bigg|\sum^{n-1}_{i=0} \frac{c_i}{x-\alpha_i}\equiv 0 \pmod {g(x)}\right\}\]

In other words, the binary Goppa code $C(g,L)$ consists of the set of all vectors that belong to the kernel of the syndrome function \cite{Barreto2012} \[s(x)\equiv \sum^{n-1}_{i=0} \frac{c_i}{x-\alpha_i} \pmod {g(x)}\]

The parity-check matrix of a Goppa code is the $(n-k)\times n$ matrix $H=XYZ$ where
\[
X=
\begin{bmatrix}
    g_{t} & 0 & 0 & \cdots  & 0\\
    g_{t-1} & g_{t} & 0 & \cdots & 0\\
     g_{t-2} & g_{t-1} & g_t & \ddots & \vdots
    \\
    \vdots & \vdots & \vdots & \ddots & 0\\
    g_1 & g_2 & g_3 & \cdots & g_{t}
\end{bmatrix}
,\;
Y=\begin{bmatrix}
    1 & 1 & \cdots & 1\\
    \alpha^{1}_0 & \alpha^{1}_1 & \cdots & \alpha^{1}_{n-1}\\
    \vdots & \vdots & \ddots & \vdots\\
    \alpha^{t-1}_0 & \alpha^{t-1}_1 & \cdots & \alpha^{t-1}_{n-1}
\end{bmatrix}
\mathrm{, \; and}\]
\[
Z=\begin{bmatrix}
    \frac{1}{g(\alpha_0)} & 0 & 0 & \cdots & 0\\
    0 & \frac{1}{g(\alpha_1)} & 0 & \cdots & 0\\
    0 & 0 & \frac{1}{g(\alpha_2)} & \ddots & \vdots
    \\
    \vdots & \vdots & \ddots & \ddots & 0\\
   0 & 0 & \cdots & 0 & \frac{1}{g(\alpha_{n-1})}
\end{bmatrix}
\]
Given that $GH^T=0$, the $k\times n$ generator matrix $G$ of a binary Goppa code can be obtained by finding the null space of the parity-check matrix $H$.

Note that binary Goppa codes have a length of $n=2^m$, a minimum dimension of $n-mt$, and a minimum distance of $2t+1$. Thus, the parameters of a $[n,k,d]$-Goppa code are $[2^m, \ge n-mt, \ge 2t+1]$.

\subsection{The McEliece Cryptosystem}
Proposed by Robert J. McEliece in 1978, the McEliece cryptosystem is a public-key encryption algorithm that has resisted cryptanalytic attacks to this day and serves as the basis for the majority of code-based systems. Despite being invented around the same time as RSA, the McEliece cryptosystem was not widely used due to its larger key size. However, due to the security it provides against quantum computers, the cryptosystem has received attention as a strong candidate for post-quantum cryptography.

\subsubsection{Key Generation}
Suppose Alice wants to send a message to Bob. In order for this to happen, Bob must generate two keys, a public key and a private key. The public key will be shared online and is what Alice will use to encrypt her message. Bob will keep his private key secret, which he will use to decrypt Alice's message. Note that it is infeasible for an eavesdropper to determine the private key solely from the public key.

First, Bob selects a random irreducible Goppa polynomial of degree $t$ and computes its corresponding generator matrix $G$. Then, Bob chooses an arbitrary $k\times k$ binary non-singular matrix $S$ and a random $n\times n$ permutation matrix $P$ (identity matrix with columns scrambled) \cite{TW2005}.

Bob then computes his public key, which consists of the $k\times n$ matrix $\hat{G}=S\times G\times P$ and the value of $t$. Bob keeps his private key $(S,G,P)$ secret.

\subsubsection{Encryption}
In order to encrypt a message $m$, Alice must generate a random binary error vector $e$ of length $n$ and weight less than or equal to $t$. Then, to obtain the ciphertext $c$, Alice must compute \[c\equiv m\cdot \hat{G}+e\pmod2\]

Note that `$\cdot$' denotes matrix-vector multiplication and `$\times$' denotes matrix multiplication \cite{UC2024}.

If $m$ is longer than $k$ bits, Alice must break up $m$ into multiple $k$-bit sequences, each of which can be encrypted separately and then concatenated. Note that $k$ is known since it is the number of rows of the published matrix $\hat{G}$.

\subsubsection{Decryption}
To decrypt the ciphertext $c$, Bob first uses his private key to calculate $P^{-1}$. Then, Bob computes
\[\hat{c}\equiv c\cdot P^{-1}\equiv m\cdot S\times G + e\cdot P^{-1} \pmod2\]
Next, Bob decodes $\hat{c}$ to the scrambled message $v\equiv m\cdot S \pmod 2$ using Patterson's Algorithm, an efficient decoding algorithm. Note that the permuted error vector $e\cdot P^{-1}$ has weight $t$ since $P^{-1}$ simply scrambles the elements of the error vector, which means that decoding still works \cite{TW2005}.

Finally, Bob recovers $m$ by calculating $m\equiv v\cdot S^{-1} \pmod 2$.

\subsubsection{Encryption and Decryption Example}
For the sake of simplicity, this example will use small parameters for the binary Goppa code; however, note that in practice, the McEliece cryptosystem with these parameters would not be secure.

Suppose Alice wants to send a message to Bob. First, Bob selects a $[16,8,5]$ binary Goppa code and finds its generator matrix \cite{Knospe}
\[G=\left[
    \begin{array}{*{20}c}
    1 & 0 & 0 & 0 & 0 & 0 & 0 & 0 & 1 & 0 & 0 & 1 & 0 & 1 & 0 & 1\\
    0 & 1 & 0 & 0 & 0 & 0 & 0 & 0 & 1 & 0 & 1 & 0 & 0 & 1 & 1 & 0\\
    0 & 0 & 1 & 0 & 0 & 0 & 0 & 0 & 1 & 1 & 0 & 0 & 0 & 0 & 1 & 1\\
    0 & 0 & 0 & 1 & 0 & 0 & 0 & 0 & 1 & 1 & 1 & 0 & 1 & 0 & 0 & 0\\
    0 & 0 & 0 & 0 & 1 & 0 & 0 & 0 & 0 & 1 & 0 & 1 & 1 & 1 & 0 & 0\\
    0 & 0 & 0 & 0 & 0 & 1 & 0 & 0 & 0 & 1 & 0 & 1 & 1 & 0 & 1 & 1\\
    0 & 0 & 0 & 0 & 0 & 0 & 1 & 0 & 0 & 0 & 1 & 1 & 1 & 1 & 1 & 1\\
    0 & 0 & 0 & 0 & 0 & 0 & 0 & 1 & 0 & 1 & 1 & 1 & 0 & 0 & 1 & 0\\
    \end{array}
    \right]
\]

Next, Bob randomly chooses an invertible matrix
\[S=\begin{bmatrix}
    0 & 0 & 1 & 0 & 0 & 1 & 1 & 0 \\
    1 & 0 & 1 & 0 & 1 & 0 & 1 & 0 \\
    0 & 0 & 0 & 0 & 0 & 1 & 0 & 0 \\
    1 & 1 & 0 & 0 & 0 & 0 & 0 & 0 \\
    0 & 1 & 0 & 1 & 0 & 1 & 0 & 1 \\
    1 & 1 & 1 & 1 & 1 & 0 & 0 & 0 \\
    0 & 0 & 0 & 1 & 0 & 1 & 0 & 1 \\
    0 & 0 & 1 & 1 & 1 & 1 & 1 & 0
\end{bmatrix}
\]
and a permutaion matrix
\[
P=\begin{bmatrix}
     e_4,\; e_5 ,\; e_1 ,\; e_{12} ,\; e_{13} ,\; e_{11} ,\; e_{16} ,\; e_2 ,\; e_8 ,\; e_9 ,\; e_{15} ,\; e_7 ,\; e_6 ,\; e_3 ,\; e_{10} ,\; e_{14}
\end{bmatrix}
\]
where the columns are the standard basis vectors $e_i$.

Bob then computes his public key
\[\hat{G}=S\times G\times P=\left[\begin{array}{*{16}c}
    0 & 0 & 0 & 0 & 0 & 1 & 1 & 0 & 0 & 1 & 1 & 1 & 1 & 1 & 0 & 1 \\
    0 & 1 & 1 & 1 & 0 & 1 & 1 & 0 & 0 & 0 & 0 & 1 & 0 & 1 & 0 & 1 \\
    0 & 0 & 0 & 1 & 1 & 0 & 1 & 0 & 0 & 0 & 1 & 0 & 1 & 0 & 1 & 0 \\
    0 & 0 & 1 & 1 & 0 & 1 & 1 & 1 & 0 & 0 & 1 & 0 & 0 & 0 & 0 & 0 \\
    1 & 0 & 0 & 0 & 0 & 1 & 1 & 1 & 1 & 0 & 1 & 0 & 1 & 0 & 1 & 1 \\
    1 & 1 & 1 & 0 & 0 & 0 & 0 & 1 & 0 & 0 & 0 & 0 & 0 & 1 & 1 & 1 \\
    1 & 0 & 0 & 0 & 0 & 0 & 1 & 0 & 1 & 1 & 0 & 0 & 1 & 0 & 1 & 0 \\
    1 & 1 & 0 & 1 & 0 & 0 & 1 & 0 & 0 & 0 & 1 & 1 & 1 & 1 & 0 & 0 
    \end{array}\right]
\]
and makes his private key $(S,G,P)$.

Alice wants to encrypt the message $m=(1,0,1,0,0,1,1,0)$, so she chooses an error vector $e=(0,0,0,0,0,1,0,0,0,0,0,1,0,0,0,0)$ of weight 2 and computes
\[c\equiv m\cdot \hat{G}+e\equiv 
(0,1,1,1,1,0,1,1,1,0,0,0,1,0,1,0)\pmod 2
\]
which she sends to Bob.

Bob then calculates
\[\hat{c}\equiv c\cdot P^{-1}\equiv m\cdot S\times G + e\cdot P^{-1}\equiv (1,1,0,0,1,0,0,1,1,1,1,1,0,1,0,0) \pmod2\]
Next, Bob uses the Patterson algorithm to decode $\hat{c}$ to the message $v=m\cdot S\equiv (1,1,0,0,1,1,1,1) \pmod 2$

Finally, Bob recovers $m$ by calculating 
\[m\equiv v\cdot S^{-1}\equiv (1,0,1,0,0,1,1,0) \pmod 2\]
\subsubsection{Security}
Suppose a third party named Eve intercepts the encrypted message sent from Alice to Bob. Eve has access to the public information, so she knows $\hat{G}$. Since she intercepts the ciphertext, she also knows the value of $c=m\cdot \hat{G}+e$, as well as the weight $t$ of the error vector $e$.

Without the private key, there are generally two attacks that Eve would attempt. First, Eve may try to determine the structure of the original code from a generator matrix of an equivalent code (i.e. recover $G$ from $\hat{G}$). Note that two linear codes $C$ and $\hat{C}$ are equivalent if there exists an invertible matrix $S$ and a permutation matrix $P$ such that $G_{\hat{C}}=S\times G_{C}\times P$. This explains why $G$ and $\hat{G}=S\times G\times P$ generate equivalent codes. However, with suitable parameter choices for Goppa codes, the generator matrix of the equivalent code does not present any visible structure that an attacker could exploit (i.e. it is difficult to distinguish between a permuted Goppa code and a random linear code), meaning that it is infeasible to determine $G$ from $\hat{G}$ \cite{Sendrier1998}. Also, note that without knowledge of the matrices $S$ or $P$, determining $G$ by calculating $S^{-1}\times\hat{G}\times P^{-1}=G$ is infeasible given the immense amount of possibilities for the matrices $S$ and $P$ when $n$ and $t$ are large \cite{McEliece78}.

Secondly, Eve may attempt to determine the message $m$ directly from the ciphertext $c$ without finding $G$. In other words, Eve is attempting to decode a random linear code with $t$ errors, which is proven to be NP-hard \cite{DecodingProblem}.

In his initial paper, McEliece proposed the use of the parameters $[n,k,t]=[1024, 524, 50]$ because these parameters result in $10^{149}$ potential Goppa polynomials and an immense amount of possibilities for $S$ and $P$ \cite{McEliece78}. McEliece noted that a brute-force approach based on decoding each codeword has a work factor of $2^k=2^{524}$ (i.e. $2^{524}$ iterations performed), which is infeasible \cite{McEliece78}. Moreover, a brute-force approach based on coset leaders has a work factor of $2^{n-k}=2^{500}$, which is also infeasible \cite{McEliece78}.

However, in 2008, improvements were made to Stern's attack on the McEliece cryptosystem, which made it possible to break the system with parameters $[1024,524,50]$ in $2^{60.55}$ bit operations \cite{McElieceNKT}. Thus, the parameters were revised to $[2048, 1751, 27]$, which guaranteed $80$-bit security against the attack \cite{McElieceNKT}.

For security against attacks from quantum computers, larger parameters are certainly needed. In fact, for $128$-bit post-quantum security, the parameters $[6960,5413,119]$ are recommended \cite{PQCMcEliece}.

\subsubsection{Key Size}
The public-key of the McEliece cryptosystem is the $k\times n$ matrix $\hat{G}=S\times G\times P$, meaning that the key size is $k\times n$ bits. However, the matrix $S$ is typically chosen so that it transforms the matrix $\hat{G}$ into systematic form, meaning the last $k$ columns of $\hat{G}$ consist of the $k\times k$ identity matrix $I$. Thus, only the first $n-k$ columns have to be stored, and therefore, the key size can be reduced to $k\times (n-k)$ bits.

Thus, the aforementioned parameters have the following key sizes:

\begin{tabularx}{0.8\textwidth} { 
  | >{\raggedright\arraybackslash}X 
  | >{\centering\arraybackslash}X 
  | >{\raggedleft\arraybackslash}X | }
 \hline
 \textbf{Parameters} & \textbf{Key Size} \\
 \hline
 [1024,524,50]  & 262,000 \\
\hline
[2048, 1751, 27]  & 520,047 \\
\hline
[6960,5413,119]  & 8,373,911\\
\hline
\end{tabularx}
\\
\\
These key sizes are substantially larger than other cryptographic methods. For instance, RSA can withstand attacks from classical computers with a key size of only 2,048 bits. Yet, there are currently efforts to reduce the key size of the McEliece cryptosystem without compromising the security it provides against attacks from quantum computers.

\subsection{Variations of the McEliece Cryptosystem}
The original McEliece cryptosystem is the basis for code-based cryptographic systems. For example, Harald Niederreiter developed the Niederreiter cryptosystem in 1986, which initially proposed using Generalized Reed-Solomon (GRS) codes before being proved insecure in 1992 \cite{Niederreiter}. However, when using Goppa codes, the Niederreiter cryptosystem reaches equivalent security to the McEliece cryptosystem, while also being faster. It functions in a similar way to McEliece's proposal but uses a parity-check matrix $H$ rather than a generator matrix $G$. 

Moreover, Classic McEliece, a key encapsulation method (KEM) based on Niederreiter's system with Goppa codes, is a finalist in NIST's Post-Quantum Cryptography Standardization competition. A KEM randomly generates a public key and a private key. A sender then uses the public key to encrypt a shared secret key, which the recipient will then decrypt with their private key. This transmitted key will then be used in a symmetric key cryptosystem, which is typically more efficient than public-key cryptosystems.

\section{Lattice-Based Cryptography}
Lattice-based cryptographic systems such as NTRU are based on lattices, and their security is dependent on the difficulty of solving problems like the Shortest and Closest Vector Problems, which become increasingly complex in higher dimensions.

\subsection{Lattices}
A lattice is a discrete subgroup of $\mathbb{R}^n$ that geometrically appears like a periodic set of points, such as the two-dimensional lattice displayed in Figure 1 below:

\begin{tikzpicture}
    \def\xmax{3}
    \def\ymax{3}
    \def\xmin{-3}
    \def\ymin{-3}
    \foreach \x in {\xmin,...,\xmax} {
        \foreach \y in {\ymin,...,\ymax} {
            \fill (\x,\y) circle (2pt);
        }
    }
    \def\axismin{-3}
    \def\axismax{3}
    \draw[thin,<->] (\axismin-1,0) -- (\axismax+1,0) node[right] {$x$};
    \draw[thin,<->] (0,\axismin-0.5) -- (0,\axismax+0.5) node[above] {$y$};
    \draw[blue, ultra thick,->] (0,0) -- (1,0) node[right] {$e_1$};
    \draw[blue, ultra thick,->] (0,0) -- (0,1) node[above] {$e_2$};
    \foreach \x in {\axismin,...,\axismax} {
        \node at (\x,-0.3) {\x};
    }
    \foreach \y in {\axismin,...,\axismax} {
        \node at (-0.3,\y) {\y};
    }
    \node at (0,-0.3) {0};
    \node [below=3.5cm, align=flush center,text width=8cm] 
        {
            \textbf{Figure 1}
        };
\end{tikzpicture}

Every lattice is generated by a basis. For a linearly independent set of vectors $B=\{b_0,...,b_{n-1}\}$, the lattice generated by $B$ is 
\[\Lambda(B)=\left\{x_i\in \mathbb{Z}\Bigg|\sum^{n-1}_{i=0} x_i\cdot b_i\right\}\]
In other words, the lattice is generated by taking integer linear combinations of the basis vectors. For instance, the lattice in Figure 1 could be generated by the basis $B=\{e_1,e_2\}=\{(1,0), (0,1)\}$.

The basis of a lattice is not unique and in fact, a lattice has an infinite number of bases. Note that a ``good" basis for a lattice consists of short, almost orthogonal vectors, whereas a ``bad" basis consists of long vectors with a small angle between them \cite{Regev}. As depicted below in Figure 2, the basis $B_1=\{(1,0),(0,1)\}$ is a good basis for the lattice, whereas the basis $B_2=\{(3,0), (5,1)\}$ is a bad basis.

\begin{tikzpicture}
    % Define lattice dimensions
    \def\xmax{5}
    \def\ymax{5}

    % Draw the lattice points
    \foreach \x in {0,1,...,\xmax} {
        \foreach \y in {0,1,...,\ymax} {
            \fill (\x,\y) circle (2pt);
        }
    }

    % Draw x and y axes
    \draw[thick,->] (-0.5,0) -- (\xmax+0.5,0) node[right] {$x$};
    \draw[thick,->] (0,-0.5) -- (0,\ymax+0.5) node[above] {$y$};

    \draw[blue, ultra thick,->] (0,0) -- (1,0) node[right] {$(1,0)$};
    \draw[blue, ultra thick,->] (0,0) -- (0,1) node[above] {$(0,1)$};

    \draw[red, ultra thick,->] (0,0) -- (3,0) node[right] {$(3,0)$};
    \draw[red, ultra thick,->] (0,0) -- (5,1) node[above] {$(5,1)$};

    % Add axis labels
    \foreach \x in {1,2,...,\xmax} {
        \node at (\x,-0.3) {\x};
    }
    \foreach \y in {1,2,...,\ymax} {
        \node at (-0.3,\y) {\y};
    }

    \node at (3,1)[below=1.6cm, align= flush center,text width=8cm] 
        {
            \textbf{Figure 2}
        };
\end{tikzpicture}

The security of lattice-based cryptographic systems is typically based on three problems that are presumed to be computationally difficult \cite{Repka}: 
\begin{enumerate}
\item Shortest Vector Problem (SVP): Given a random ``bad" basis $B$ for a lattice $\Lambda$, find the shortest nonzero vector in $\Lambda$.
\item Closest Vector Problem (CVP): Given an arbitrary ``bad" basis $B$ for a lattice $\Lambda$ and a target point $x$, find the lattice point closest to $x$.
\item Shortest Independent Vectors Problem (SIVP): Given an arbitrary basis $B\in\mathbb{Z}^{n\times n}$ for a lattice $\Lambda$, find a new basis $\hat{B}=\{\hat{b}_0,...,\hat{b}_{n-1}\}$ for $\Lambda$ that minimizes the value of $\max_i||\hat{b}_i||$
(i.e. minimizes the length of the longest vector in the basis).
\end{enumerate}

The primary approach to these problems involves basis reduction, in which an arbitrary lattice basis is transformed into a new basis that consists of shorter, nearly orthogonal vectors. In fact, one of the most successful algorithms, the Lenstra-Lenstra-Lovász (LLL) algorithm, can approximate a reduced lattice basis in polynomial time \cite{LLL1982}. However, it is important to note that the LLL algorithm is not effective in higher dimensions n such as $n\geq 100$ because the quality of the approximation decreases (i.e. the reduced basis vectors are not as short or orthogonal) \cite{TW2005}. Currently, there is no known polynomial-time algorithm that can solve these problems with classical or quantum computers, which has made lattice-based cryptographic systems ideal candidates for post-quantum cryptography.

\subsection{NTRU Cryptography}
NTRU is a lattice-based, asymmetric cryptographic system that is a finalist in NIST's Post-Quantum Cryptography Standardization project. Its security is based on the Shortest Vector Problem, which becomes computationally difficult in higher dimensions. As a result, NTRU has been resistant to attacks from classical and quantum computers, which makes it an excellent candidate for post-quantum cryptography.

\subsubsection{Arithmetic in NTRU}
NTRU is defined over the polynomial ring $R=\mathbb{Z}$[X]/($X^N-1)$, meaning the system reduces polynomials with integer coefficients modulo $x^N-1$.

For a ring R, the following properties are satisfied \cite{UC2024}:
\begin{enumerate}
\item Closed Under Addition and Multiplication: $\forall a,b\in R$, $(a+b)\in R$ and $(a\cdot b)\in R$
\item Associative Property of Addition and Multiplication: $\forall a,b,c\in R$, $(a+b)+c=a+(b+c)$ and $(a\cdot b)\cdot c=a\cdot (b\cdot c)$
\item Commutative Property of Addition: $\forall a,b\in R$, $a+b=b+a$
\item Additive Identity: $\forall a\in R$, $a+0=a$
\item Additive Inverse: $\forall a\in R$, there exists the negative element $-a$ such that $a+(-a)=0$
\item Multiplicative Identity: $\forall a\in R$, $a\cdot 1=a$
\item Distributive Property: $\forall a,b,c\in R$, $a\cdot (b+c)=(a\cdot b) + (a\cdot c)$
\end{enumerate}
In a polynomial ring, addition is the same as traditional polynomial addition, where the coefficients of ``like terms" (terms with the same degree) are added together. Multiplication is defined by the cyclic convolution product \cite{Hoffstein}:
\[f\circlesign{*}g=h \hspace{0.15 cm} \mathrm{with} \hspace{0.15 cm} h_k=\sum^k_{i=0}f_ig_{k-i} + \sum^{N-1}_{i=k+1}f_ig_{N+k-i}=\sum_{i+j\equiv k \pmod N}f_ig_j\]
and is denoted by $\circlesign{*}$. This is the same as multiplying the polynomials out and reducing modulo $x^N-1$. When performing multiplication modulo an integer $q$, reduce the coefficients modulo $q$ and the polynomial modulo $x^N-1$.

\subsubsection{Key Generation}
Suppose Alice wants to send a message to Bob. First, Bob must select integer values for the parameters of the system, $N$, $p$, and $q$. Note that $p$ and $q$ must be relatively prime (gcd$(p,q)=1$) and that $p$ must be smaller than $q$ \cite{TW2005}.

Next, Bob must arbitrarily generate two polynomials, $f$ and $g$, with degrees less than $N$ and small coefficients, typically between $-1$ and $1$. Also, $f$ must be invertible modulo $p$ and modulo $q$, meaning there exist $f^{-1}_p$ and $f^{-1}_q$ such that $f^{-1}_p * f\equiv 1 \pmod p$ and $f^{-1}_q * f\equiv 1 \pmod q$ \cite{TW2005}.

Then, Bob efficiently calculates the inverses $f^{-1}_p$ and $f^{-1}_q$ using the extended Euclidean algorithm. Bob computes $h\equiv f^{-1}_q\circlesign{*} g \pmod q$ and makes his public key ($N, p, q, h$) \cite{TW2005}. Bob keeps ($f, f^{-1}_p$) as his private key and discards $g$, as it will no longer be used \cite{TW2005}. Note that it is not necessary to store $f^{-1}_p$ since it can be efficiently calculated from $f$, but storing the value makes decryption faster.

\subsubsection{Encryption}
In order to encrypt her message, Alice must first convert it into a polynomial $m$ with degree less than $N$ and coefficients modulo $p$. Note that in this case, coefficients modulo $p$ refers to the integers from $-\frac{p}{2}$ to $\frac{p}{2}$. For instance, if $p=3$, the coefficients range from $-1$ to $1$.

Then, Alice must choose an arbitrary small polynomial $r$ and compute the ciphertext \cite{Hoffstein}
\[c\equiv pr\circlesign{*}h + m \pmod q\]
which she sends to Bob. Note that reduction modulo $q$ refers to reduction of the coefficients to the integers in the interval ($-\frac{q}{2},\frac{q}{2}$], not the interval [$0,q-1$].

\subsubsection{Decryption}
To decrypt the ciphertext c, Bob must first calculate \cite{Hoffstein}
\[a\equiv f\circlesign{*}c \pmod q\]
with integer coefficients from ($-\frac{q}{2},\frac{q}{2}$].

Then, Bob can recover the original message by calculating \cite{TW2005}
\[m\equiv f^{-1}_p \circlesign{*} a \pmod p\]
with integer coefficients from ($-\frac{p}{2},\frac{p}{2}$].

\subsubsection{Encryption and Decryption Example}
This example will use smaller parameters for the sake of simplicity; however, note that in practice, the NTRU cryptosystem with these parameters would not be secure.

Suppose Alice wants to send a message to Bob. First, Bob selects $N=11$, $p=3$, and $q=41$ as the parameters for the system. Next, he chooses two polynomials $f$ and $g$ that have coefficients in the range $[-1,1]$:
\[f=x^9-x^8+x^6+x^5-x+1 \hspace{0.15cm} \mathrm{and} \hspace{0.15cm} g=x^{10}+x^7-x^6+x^3+x^2-1\]
Then, Bob calculates the inverses \[f^{-1}_p=2x^8+x^7+x^6+x^5+2x^2+x\] and \[f^{-1}_q=24x^{10}+7x^8+17x^7+8x^6+35x^5+17x^4+32x^3+25x^2+31x+30\]
Next, Bob calculates
\[h\equiv f^{-1}_q\circlesign{*}g\equiv 4x^{10}+x^9+x^8+11x^6+6x^5+40x^4+15x^3+11x^2+4x+31\pmod {41}\]
and makes his public key $(N,p,q,h)$.

To encrypt her message $m=x^9+x^5-x^3+1$, Alice chooses a random polynomial $r=x^4+x^2-1$ and computes the ciphertext
\[c\equiv pr\circlesign{*}h + m\equiv -17x^{10}+16x^9-14x^8-19x^7-3x^6-x^5+6x^4+19x^3-19x^2+3x-7  \pmod {41}\]
Note that Alice must reduce coefficients to the integers within the interval (-$\frac{41}{2}$, $\frac{41}{2}$] before sending the ciphertext $c$ to Bob.

To decrypt the ciphertext $c$, Bob first computes
\[a\equiv f\circlesign{*}c\equiv -6x^{10}+4x^9-5x^8+x^7+5x^6+5x^5+2x^4+x^3-7x^2+x+9  \pmod {41}\]
reducing coefficients to the integers from the interval (-$\frac{41}{2}$, $\frac{41}{2}$].

Finally, Bob retrieves the original message by calculating
\[m\equiv f^{-1}_p\circlesign{*}a\equiv x^9+x^5-x^3+1 \pmod 3\]
and reducing coefficients to the integers within the interval (-$\frac{3}{2}$, $\frac{3}{2}$].
\subsubsection{Why Decryption Works}
Bob calculates the polynomial
\begin{eqnarray*}
a&\equiv& f\circlesign{*}c\equiv f\circlesign{*}(pr\circlesign{*}h + m) \pmod q 
\\
&\equiv& pr\circlesign{*}f\circlesign{*}f^{-1}_q \circlesign{*} g + f\circlesign{*}m \pmod q
\\
&\equiv& pr \circlesign{*} g + f \circlesign{*} m \pmod q
\end{eqnarray*}
Recall that $p$ is smaller than $q$ and that the coefficients for $r$, $g$, $f$, and $m$ are small, meaning that it is likely that before reduction modulo $q$, the polynomial $a$ already has integer coefficients from the interval ($-\frac{q}{2},\frac{q}{2}$] \cite{TW2005}.

Therefore,
\begin{eqnarray*}
f^{-1}_p\circlesign{*} a&\equiv& f^{-1}_p\circlesign{*}(pr \circlesign{*} g) + f^{-1}_p\circlesign{*}(f \circlesign{*}m) \pmod p
\\
&\equiv& p*(f^{-1}_p\circlesign{*}r \circlesign{*} g) + 1*m \pmod p
\\
&\equiv& m \pmod p
\end{eqnarray*}
meaning that decryption works.

\subsubsection{Connection to a Lattice}
Although NTRU does not initially seem related to lattices and the Shortest Vector Problem, the process can be equivalently expressed in terms of lattices.

\textbf{Terminology}: Similar to the cyclic shift $\pi$ in linear codes, let $L$ be a linear transformation that cyclically shifts the components of a vector \cite{Regev}. Define $L*v=[v, Lv, L^2v,..., L^{N-1}v]$ to be the circulant matrix of a vector $v$ of length $N$ (i.e. the matrix with rows consisting of the cyclic shifts of $v$) \cite{Regev}. NTRU uses convolutional modular lattices. A convolutional modular lattice $\Lambda(c,q)$ is the set of vectors $\Lambda(c,q)=\{(a,b)\in \mathbb{Z}^{2N}:a\circlesign{*}c\equiv b \pmod q\}$ \cite{May}.

\textbf{Key Generation}: 
The parameters of the system are a prime number $N$, a small integer $p$, an integer modulus $q$, and an integer weight bound $d_f$ \cite{Regev}. Instead of two polynomials, the private key will be the short vector $(f,g)=(f_0,...,f_{N-1},g_0,...,g_{N-1})\in \mathbb{Z}^{2N}$. Note that $f$ and $g$ must be chosen such that
\begin{enumerate}
    \item $[L*f]$ is invertible modulo $q$
    \item $f\in e_1 + \{-p,0,p\}^N$ ($e_1$ refers to the standard basis vector) and $g\in \{-p,0,p\}^N$ where $f-e_1$ and $g$ have $d_f+1$ positive entries and $d_f$ negative entries. Note that as a result of this restriction, $[L*f]\equiv I \pmod p$ and $[L*g]\equiv 0 \pmod p$.
\end{enumerate}
The associated lattice to $(f,g)$ is $\Lambda((L*f,L*g)^T,q)$, which is the smallest convolutional lattice that contains $(f,g)$. Note that when $N$ is sufficiently large, there is a high probability that $(f,g)$ is the shortest vector in the lattice \cite{Shaw}.

The public key is chosen to be the Hermite Normal Form basis of the lattice $\Lambda((L*f,L*g)^T,q)$, which is considered to be a ``bad" basis for the lattice \cite{Regev}. Thus, the public key is
\[H=
\begin{bmatrix}
    I & 0\\
    L*h & q\cdot I\\
\end{bmatrix}
\]
where $h\equiv[L*f]^{-1}g \pmod q$ \cite{Regev}.

\textbf{Encryption}: A message is expressed as a vector $m\in \{-1,0,1\}^N$ that has exactly $d_f+1$ positive components and $d_f$ negative entries. To encrypt the message $m$, it is concatenated with an arbitrary error vector $r\in \{-1,0,1\}^N$ that has $d_f+1$ positive components and $d_f$ negative entries. Then, this resulting vector is reduced modulo $H$ \cite{Regev}:
\[\begin{bmatrix}
    r \\
    m \\
\end{bmatrix}
\bmod \begin{bmatrix}
    I & 0\\
    L*h & q\cdot I\\
\end{bmatrix}
=
\begin{bmatrix}
    0\\
    (m+[L*h]r) \bmod q\\
\end{bmatrix}
\]
Given that the first $N$ components of the vector are always 0, the ciphertext is the $N$-bit vector \[c\equiv(m+[L*h]r) \pmod q\]

\textbf{Decryption}: To decrypt the ciphertext $c$, compute 
\[[L*f]c\equiv[L*f]m + [L*f][L*h]r\equiv [L*f]m + [L*g]r\pmod q\]
Since the vector $[L*f]m + [L*g]r$ already has components between $-\frac{q}{2}$ and $\frac{q}{2}$, we can reduce the vector modulo $p$ to obtain $m$:
\[[L*f]m + [L*g]r\equiv I\cdot m+0\cdot r\equiv m \pmod p\]

\subsubsection{Security}
Suppose a third party named Eve intercepts the encrypted message sent from
Alice to Bob.
Eve then has access to the ciphertext $c\equiv(m+[L*h]r) \pmod q$, as well as the public key $H=\begin{bmatrix}
    I & 0\\
    L*h & q\cdot I\\
\end{bmatrix}$.

The public key $H$ provides a bad basis for the lattice $\Lambda((L*f,L*g)^T,q)$. Given that there is a high probability that $(f,g)$ is the shortest vector in the lattice, determining the private key requires Eve to solve the Shortest Vector Problem, which is infeasible. Eve could attempt basis reduction with an algorithm such as the LLL algorithm; however, it has previously been stated that this is not effective in higher dimensions.

For optimal security, it is recommended that $p=3$ and that $q$ is a power of 2. To achieve 128-bit post-quantum security, the parameters $[N,p,q]=[443,3,2048]$ are recommended \cite{NTRUParameters}. Note that $N$ could also be $587$ or $743$.

\section{Connections Between the Systems}
Code-based and lattice-based cryptography initially appear very different, with one process revolving around linear codes and the other around a periodic structure from abstract algebra. However, when compared more closely, these processes actually have many similarities.

\subsection{Linear Codes and Lattices}
Despite being distinct, linear codes and lattices are very similar at their core. In fact, one major similarity is that both linear codes and lattices are generated by bases.

Note that a set of vectors $B$ is considered a basis for a vector space $V$ if: \begin{enumerate}
    \item $B$ is a linearly independent set of vectors
    \item $B$ spans $V$ (i.e. every vector in $V$ can be obtained by taking linear combinations of the basis vectors)
\end{enumerate}

A linear code $C$ is a code that is closed under addition, meaning that $\forall u,v\in C$, $u+v\in C$. Linear codes have various bases. In fact, a linear code $C$ of dimension $k$ has $\frac{1}{k!}\prod_{i=0}^{k-1}(2^k-2^i)$ different bases \cite{Hankerson}.

For example, a basis for the vector space $K^3$ is $B=\{100,010,001\}$, given that $B$ is a linearly independent set of words that spans $K^3$. In contrast, despite being a linearly independent set of vectors, $\hat{B}=\{100,010\}$ is not a basis for $K^3$ because it does not span the space. For instance, the word $001$ cannot be produced by any linear combination of the basis vectors.

As previously mentioned, lattices are also described by their infinite bases and are generated by taking integer linear combinations of the basis vectors.

\subsection{Code Equivalence and Basis Reduction}
The security of various code-based and lattice-based cryptographic systems have respectively been based on the intractability of code equivalence problems and basis reduction. In fact, the code equivalence problem in the McEliece cryptosystem and the problem of basis reduction in the NTRU cryptosystem have similar cryptographic functions.

In the McEliece cryptosystem, the difficulty of the code equivalence problem is incorporated in the public key. The generator matrix $G$ for the binary Goppa code is scrambled by two matrices, $S$ and $P$, to create the public key $\hat{G}$, which generates an equivalent code to $G$. This new matrix $\hat{G}$ does not present any structure that an attacker could exploit, meaning that it is infeasible to determine $G$ from the equivalent code $\hat{G}$. In this sense, code equivalence is used as a mask for the private key ($S$, $G$, $P$). Note that although the code equivalence problem is presumed to be difficult, it has been proven that the code-equivalence problem is not NP-complete \cite{Petrank}. 

In the lattice description of the NTRU cryptosystem, the public key is a bad basis for the lattice. Given that there is no efficient algorithm for basis reduction in higher dimensions, it is infeasible to determine a good basis for the lattice and therefore, find the shortest vector in it. Thus, similar to code equivalence, the difficulty of basis reduction helps conceal the private key.

\subsection{Minimum Distance and Shortest Vector Problems}
The Minimum Distance Problem (MDP) in code-based cryptography is very similar to the Shortest Vector Problem (SVP) in lattice-based cryptography.

In MDP, given a linear code $C$ over a finite field, one must find the non-zero codeword of minimum Hamming weight (codeword with the least number of 1's). Note that finding the minimum distance of a binary linear code is NP-hard \cite{Vardy}.

In SVP, given a bad basis $B$ for a lattice $\Lambda$, one must determine the shortest nonzero vector in $\Lambda$. Yet, without an efficient algorithm to perform basis reduction, solving SVP is infeasible.

Both the Minimum Distance Problem and Shortest Vector Problem require one to find the smallest vector in a given space. SVP wants the shortest vector in terms of Euclidean distance to the origin, whereas MDP wants the smallest vector in terms of Hamming distance to the zero word. Their shared computational difficulty makes them ideal problems to use in cryptographic systems.

\subsection{Nearest Codeword and Closest Vector Problems}
The Nearest Codeword Problem (NCP) in code-based cryptography closely resembles the Closest Vector Problem (CVP) in lattice-based cryptography.

In NCP, given a linear code $C$ and a word $w$, one must find the codeword $v$ closest to $w$ (minimum Hamming distance), which is NP-hard \cite{DecodingProblem}.

In CVP, given a bad basis $B$ for a lattice $\Lambda$ and a target point $x$, one needs to find the lattice point closest to $x$, which is infeasible in higher dimensions.

Both problems require one to find the closest vector to another in a given space. While solving NCP means finding the nearest codeword in terms of Hamming distance, CVP operates in terms of Euclidean distance. Similar to MDP and SVP, the intractability of these problems makes them useful in cryptography.

The connections between these two problems are evident in the structure of some modern, polynomial-time decoding algorithms for irreducible binary Goppa codes. A list-decoding algorithm, as well as an altered version of the original Patterson's Algorithm, implement lattice basis reduction in the decoding process \cite{ListDecoding}. Thus, when performing decoding in code-based systems such as McEliece, basis reduction, which is typically found in lattice-based cryptographic systems, is employed, which further displays the similarities between these two systems.

\subsection{Security Comparison}
Now, let's compare the security of the McEliece cryptosystem and NTRU.

The Nearest Codeword Problem, which is the basis of the McEliece cryptosystem, is simply the Closest Vector Problem defined in terms of Hamming distance, which makes the two problems equivalent. Since the Closest Vector Problem is presumed to be more difficult than the Shortest Vector Problem, which is the basis of NTRU, the McEliece cryptosystem could be considered more secure than NTRU.

However, if an efficient algorithm is found to distinguish between binary Goppa codes and random binary linear codes, the McEliece cryptosystem could be cracked because the public key would leak information to an eavesdropper. NTRU would remain intact, which makes it more secure on this front.

Yet, if an efficient algorithm is discovered to solve the Closest Vector Problem, the security of both cryptosystems would be compromised. 

Thus, given that there are multiple ways to attack each system, it is difficult to definitively compare their security.

\section{Conclusion}
With the threat that quantum computers will soon pose to our encrypted data, it is imperative to consider alternative public-key encryption methods that are secure against attacks from quantum computers. 

Having resisted decades of cryptanalytic attacks from both classical and quantum computers, the security of code-based systems such as the McEliece cryptosystem make them good candidates for post-quantum cryptography. Moreover, with ongoing efforts to reduce the key size of the McEliece cryptosystem, it could become crucial to the future security of our digital information.

Lattice-based systems such as NTRU are arguably the most promising post-quantum cryptographic systems, with the majority of candidates in NIST's Post-Quantum Cryptography Standardization competition being lattice-based. These systems are presumed to be secure given the difficulty of problems such as SVP and CVP, and given that they typically have smaller key sizes than the McEliece scheme, lattice-based systems may be the future.

Despite initially appearing distinct, code-based and lattice-based systems have various similarities, given the strong connection between linear codes and lattices. As a result, the security of both systems seems relatively interconnected, meaning that if an efficient algorithm can break one system, it may also break the other.


\begin{thebibliography}{9}
\bibitem{Shor}
    Shor, Peter W. ``Algorithms for quantum computation: discrete logarithms and factoring." In \textit{Proceedings 35th annual symposium on foundations of computer science}, pp. 124-134. Ieee, 1994.

\bibitem{UC2024}
    Paar, Christof, Jan Pelzl, and Tim Güneysu.
    \textit{Understanding Cryptography: From Established Symmetric and Asymmetric Ciphers to Post-Quantum Algorithms}. Springer, 2024.

\bibitem{Patterson}
    Patterson, Nicholas. ``The algebraic decoding of Goppa codes." \textit{IEEE Transactions on Information Theory} 21, no. 2 (1975): 203-207.

\bibitem{Repka}
    Repka, Marek, and Pavol Zajac. ``Overview of the McEliece cryptosystem and its security." \textit{Tatra Mountains Mathematical Publications} 60, no. 1 (2014): 57-83.

\bibitem{Barreto2012}
    Barreto, Paulo SLM, Rafael Misoczki, and Richard Lindner. ``Decoding Square-Free Goppa Codes Over $\mathbb{F}_{p}$." \textit{IEEE transactions on information theory} 59, no. 10 (2013): 6851-6858. 
    
\bibitem{TW2005}
    Trappe, Wade, and Lawrence Washington.
    \textit{Intro to Cryptography with Coding Theory}. 2nd ed. Upper Saddle River: Prentice Hall, 2006.

\bibitem{Knospe}
    Knospe, Heiko. \textit{A Course in Cryptography}. Vol. 40. American Mathematical Society, 2019.

\bibitem{Sendrier1998}
    Canteaut, Anne, and Nicolas Sendrier. ``Cryptanalysis of the original McEliece cryptosystem." In \textit{Advances in Cryptology—ASIACRYPT’98: International Conference on the Theory and Application of Cryptology and Information Security Beijing, China, October 18–22, 1998 Proceedings}, pp. 187-199. Springer Berlin Heidelberg, 1998.
    
\bibitem{McEliece78}
  McEliece, Robert J.
  ``A Public-Key Cryptosystem Based on Algebraic Coding Theory." In \textit{DSN Progress Report 42–44}: 114-116 (1978).

\bibitem{DecodingProblem}
  McEliece, R., and H. Van Tilborg. ``On the inherent intractability of certain coding problems." \textit{IEEE Transactions on Information Theory} 24, no. 3 (1978): 384-386.

\bibitem{McElieceNKT}
    Bernstein, Daniel J., Tanja Lange, and Christiane Peters. ``Attacking and defending the McEliece cryptosystem." In \textit{Post-Quantum Cryptography: Second International Workshop, PQCrypto 2008 Cincinnati, OH, USA, October 17-19, 2008 Proceedings 2}, pp. 31-46. Springer Berlin Heidelberg, 2008.

\bibitem{PQCMcEliece}
    
    Augot, Daniel, Lejla Batina, Daniel J. Bernstein, Joppe Bos, Johannes Buchmann, Wouter Castryck, Orr Dunkelman et al. ``Initial recommendations of long-term secure post-quantum systems." \textit{PQCRYPTO. EU. Horizon} 2020 (2015).

\bibitem{Niederreiter}
    Niederreiter, Harald. ``Knapsack-type cryptosystems and algebraic coding theory." \textit{Prob. Contr. Inform. Theory} 15, no. 2 (1986): 157-166.
    
\bibitem{PQC2008}
    Bernstein, Daniel J., Johannes Buchmann, and Erik Dahmen.
    \textit{Post-Quantum Cryptography}. Berlin: Springer, 2008.

\bibitem{Regev}
    Micciancio, Daniele, and Oded Regev. ``Lattice-based cryptography." In \textit{Post-quantum cryptography}, pp. 147-191. Berlin, Heidelberg: Springer Berlin Heidelberg, 2009.

\bibitem{LLL1982}
    Lenstra, Arjen K., Hendrik Willem Lenstra, and László Lovász. ``Factoring polynomials with rational coefficients." \textit{Mathematische annalen} 261 (1982): 515-534.
\bibitem{Hoffstein}
    Hoffstein, Jeffrey, Jill Pipher, and Joseph H. Silverman. ``NTRU: A ring-based public key cryptosystem." In \textit{International algorithmic number theory symposium}, pp. 267-288. Berlin, Heidelberg: Springer Berlin Heidelberg, 1998.

\bibitem{May}
    May, Alexander, and Joseph H. Silverman. ``Dimension reduction methods for convolution modular lattices." In \textit{International Cryptography and Lattices Conference}, pp. 110-125. Berlin, Heidelberg: Springer Berlin Heidelberg, 2001.

\bibitem{Shaw}
    Shaw, Hayden. ``Decrypting NTRU: An Introductory Mathematical Exploration of the NTRU Public Key Cryptosystem." PhD diss., 2023.

\bibitem{NTRUParameters}
    Hoffstein, Jeff, Jill Pipher, John M. Schanck, Joseph H. Silverman, William Whyte, and Zhenfei Zhang. ``Choosing parameters for NTRUEncrypt." In \textit{Cryptographers’ Track at the RSA Conference}, pp. 3-18. Cham: Springer International Publishing, 2017.
    
\bibitem{Hankerson}
    Hankerson, D. C., Gary Hoffman, Douglas A. Leonard, Charles C. Lindner, Kevin T. Phelps, Chris A. Rodger, and James R. Wall. \textit{Coding theory and cryptography: the essentials}. CRC Press, 2000.

\bibitem{Petrank}
    Petrank, Erez, and Ron M. Roth. ``Is code equivalence easy to decide?." \textit{IEEE Transactions on Information Theory} 43, no. 5 (1997): 1602-1604.

\bibitem{Vardy}
    Vardy, Alexander. ``The intractability of computing the minimum distance of a code." \textit{IEEE Transactions on Information Theory} 43, no. 6 (1997): 1757-1766.

\bibitem{ListDecoding}
    Bernstein, Daniel J. ``List decoding for binary Goppa codes." In \textit{International Conference on Coding and Cryptology}, pp. 62-80. Berlin, Heidelberg: Springer Berlin Heidelberg, 2011.

\end{thebibliography}
\end{document}